\documentstyle[epsfig]{mn}
\title[Pulsar Distances based on Interstellar Scattering]{Improving
Pulsar Distances by Modelling Interstellar Scattering}
\author[A.A. Deshpande \& R. Ramachandran]
{A.A. Deshpande$^1$ \& R. Ramachandran$^2$ \\ 
    $^1$Raman Research Institute, Bangalore - 560 080, India : desh@rri.ernet.in \\ 
    $^2$Sterrenkundig Instituut, Universiteit van Amsterdam, Kruislaan 403, 
        1098 SJ Amsterdam, The Netherlands : ramach@astro.uva.nl}
\begin{document}
\maketitle

\begin{abstract}
     We present here a method to study the distribution of electron 
     density fluctuations in pulsar directions as well as to estimate
     pulsar distances. The method, based on a simple two-component model 
     of the scattering medium discussed by Gwinn {\it et al.} (1993),
     uses  scintillation \& proper motion data in addition to the
     measurements of angular broadening \& temporal broadening to solve 
     for the model parameters, namely, the fractional distance to a 
     discrete scatterer and the ascociated relative scattering strength.
     We show how this method can be used to estimate pulsar distances 
     reliably, when the location of a discrete scatterer (e.g. an HII 
     region), if any, is known. Considering the specific example of PSR 
     B0736--40, we illustrate how a simple characterization of the Gum 
     nebula region (using the data on the Vela pulsar) is possible and 
     can be used along with the temporal broadening measurements to
     estimate pulsar distances.
\end{abstract}


\section{Introduction}
Reliable estimation of pulsar distances forms a crucial input for many
important investigations of pulsar properties, particularly those
concerning spatial distribution, space velocities, birth-rates etc..
The conventional method for estimation of distances is based on the measured 
value of
the column density of electrons between us and the pulsar (i.e. the dispersion
measure, DM) combined with our assumption of the distribution of 
free electrons in the Milkyway. Other methods, which give the so called 
{\it independent distance estimates},
are based on pulsar association with a supernova remnant (or a globular 
cluster), measurements
of annual parallax for nearby pulsars or the useful limits through HI
absorption measurements (possible for pulsars in the galactic plane).
These `independent' estimates provide important constraints for models 
describing the distribution of electron density in our galaxy.
 
Although the model of the electron density distribution based on pulsar data 
has received many refinements over the years
(e.g. Prentice \& ter Haar 1969; Vivekanand \& Narayan 1982; Lyne {\ et al.} 1985), 
the recent comprehensive 
model by Taylor \& Cordes (1993) represents a major qualitative improvement
wherein the spiral-arm structure has been incorporated explicitly.
This model is derived based on the HII region distribution, constraints
provided by the `independent' estimates of distances, data on scatter 
broadening of pulsar signals, the radio continuum emission associated
with our galaxy etc.. Estimation based on this model
(and using the dispersion measures of pulsars) has pushed pulsar
distances farther by a factor of 1.5 to 2 compared to earlier similar
estimates, particularly for the `local' pulsars. This has had a serious 
implication in terms of a corresponding increase in the estimated
velocities of pulsars based on the measured proper motions.

 Although this model is a considerable improvement, some features
are worth noting.
  The typical uncertainty in most of the available estimates is believed
to be about 20-30\% (rms), while in some cases, distances are uncertain by a 
factor of 2 or more. 
For example, the model is seen to 
over-estimate by a large factor (in some cases $>2$) 
the distances to pulsars at high 
galactic latitudes. An analysis of the correlation of the pulsar
distribution with the spiral arm locations (Ramachandran \& Deshpande, 1994)
points out a possible bias in the estimated pulsar locations towards
the spiral arms. This bias could be understood in terms
of a possible under-estimation of the electron density in the 
interarm regions. If this
is true, then we estimate that the use of the Taylor \& Cordes model
leads to an over-estimation of distances (using DMs) by 30\% or so 
for the local population of pulsars.
 
In light of these, the need for a more reliable distance estimator
for pulsars cannot be over-emphasized. In this paper, we explore an
attractive possibility
wherein the observables associated with the interstellar scattering can
be used in the distance estimation.

The fluctuation of the electron density in the interstellar medium gives rise
to a variation of the refractive index, which results in the scintillation 
of radio signals. The basic analysis of scintillations in terms of these 
refractive index fluctuations was presented by Scheuer (1968). Over the 
years, many authors have studied this problem in detail, and have shown 
that scattering introduces many other observable effects like apparent angular 
broadening, temporal broadening, intensity scintillations  
etc. (Scheuer 1968; Rickett 1969; Alcock \& Hatchett 1978; Goodman \& 
Narayan 1985; Blandford \& Narayan 1985). 

On the whole, the distribution of scattering material in the Galaxy can be
represented by high-density localized components associated with HII regions 
and supernova remnants, and a more diffuse uniformly distributed component. 
Gwinn {\it et al.} (1993) discuss in detail how the angular broadening and the
temporal broadening of the pulsar signal can be effectively used to study
the distribution of scattering material along the line of sight. In this
paper, we extend this idea by introducing two more observable 
parameters, the diffractive scintillation time scale $t_{\rm dif}$ and the
proper motion $\mu$, and present a method for distance estimation using the
various observables along with some possible knowledge about the distance
to the scatterer.

Sections \ref{sec-ang} \& \ref{sec-vel} present   some basic relations
that form the essence of the paper, 
connecting 
angular broadening of sources, interstellar scintillation time-scales, 
and other parameters assuming a reasonably general two-component
description for the scattering material.
In sections \ref{sec-estdist} \&
\ref{sec-probing}, we discuss how this formulation can be used to estimate
distances to pulsars, and, in turn, to study the distribution of electron 
density fluctuation in the interstellar medium. Particularly, as we
describe, this method can be used to probe and model regions
of enhanced scattering like the Gum Nebula, the Cygnus OB complex, etc..

We also discuss the specific case of PSR B0736--40, in section \ref{sec-distance},
where the recent measurement of temporal broadening (Ramachandran {\it et 
al.} 1997) has shown excess scattering attributable  to 
the Gum Nebula.
We estimate the distance to this pulsar to be $\sim$4.5 kpc, far less than the distance of $>11$ kpc derived on the 
basis of the model by Taylor \& Cordes (1993), reducing significantly the 
derived velocity of this pulsar.

\section{Apparent angular broadening}
\label{sec-ang}
The r.m.s. angular broadening of a source at a distance $D$ from the
observer is given by (Alcock \& Hatchett 1978; Blandford \& Narayan 1985)
\begin{equation}
\theta^2\;=\;\frac{1}{D^2}\int_0^D z^2 \psi(z)\;{\rm d}z
\label{eq:rmstheta}
\end{equation}
\noindent
where $z$ is the line-of-sight distance coordinate, 
whose value is zero at the location of 
the pulsar, and $D$ at the observer. $\psi(z)$ is the mean scattering rate 
per unit length. This r.m.s. broadening is related to the FWHM diameter 
$\theta_H$ of the source by $\theta_H^2 = (4\ln 2)\;\theta^2$. 

The mean temporal broadening of the pulse profile is given by (Blandford \&
Narayan 1985)
\begin{equation}
\tau_{\rm sc}\;=\;\frac{1}{2cD}\int^D_0 z\; (D-z)\;\psi(z)\; {\rm d}z
\label{eq:meandel}
\end{equation}

The mean temporal broadening is related to the decorrelation band width 
($\Delta\nu$) by the uncertainty relation: $2\pi\tau_{\rm sc}\Delta\nu = 
1$. As indicated by the above two equations, the angular broadening of the 
source is maximum when the scatterer is close to the observer, and the 
temporal broadening is maximum when the scatterer is located mid-way along
the line of sight. 
Now, let us assume that the distribution of scattering material in a given 
line of sight can be adequately described by two components: a uniformly
distributed component, and a thin screen located at a distance of $xD$ from
the observer. With this assumption, the relations in eq.s \ref{eq:rmstheta} \&
\ref{eq:meandel} can be expressed as (Gwinn {\it et al.} 1993)
\begin{eqnarray}
\theta_H^2 &=& 4\ln 2\;\frac{D\psi_0}{3} \left[ 1+3(1-x)^2 
   \frac{\psi_1}{D\psi_0}\right] \nonumber\\
\tau_{\rm sc} &=& \frac{D}{4c}\frac{D\psi_0}{3}\;\left[ 1+6x(1-x) 
   \frac{\psi_1}{D\psi_0}\right]
\label{eq:unif}
\end{eqnarray}

\noindent
Here, $\psi_1$ \& $D\psi_0$ give the mean scattering rate for the  
discrete {\it thin} screen 
and the uniform component, respectively. In practice, measuring the angular 
broadening of a source requires observations using Very-Long-Baseline 
Interferometry. In principle, it is possible to estimate 
the amount of angular broadening from the measured value of the temporal 
broadening, if we have the knowledge of the distribution of scattering 
material along the line of sight. In the absence of such knowledge,
a simple-minded estimate of the angular broadening ($\theta_{\tau}$) is 
possible and can be obtained as
\begin{equation}
\theta_{\tau}^2 = 4\ln 2\;\left(\frac{4c\tau_{\rm sc}}{D}\right) 
\end{equation}

Although such a simple-minded estimate was originally suggested for
a case of thin scatterer midway between the pulsar and the observer,
the constant in the above expression (which comes through the definition
of $\tau_{sc}$ in equation 
\ref{eq:meandel}) is such that the $\theta_{\tau}$ \& the
$\theta_H$ would match in the case of a uniformly distributed scatterer. 
With this, and given the
different dependences of the two estimates of $\theta$ on x, this
expression gives an estimate that matches $\theta_H$ even when a thin
scatterer is included at $x$=(1/3). Also note, that at $x$=(1/3), the ratio
of $\theta$s becomes independent of the relative strengths of scattering
for the two components (see Fig. 1). Thus, for consideration of the
$\theta$ ratio, a uniformly distributed scatterer can be replaced
by an equivalent thin scatterer at $x$=(1/3) and vice versa.
Using the expression for $\tau_{sc}$ and the above relation, we have  
\begin{equation}
\theta_{\tau}^2 = 4\ln 2
\frac{D\psi_0}{3}\;\left[1+6x(1-x) \frac{\psi_1}{D\psi_0} \right]
\end{equation}

Note that $\theta_{\tau}$ is not equal to $\theta_{H}$ in general. The 
difference depends on the values of $x$ and ${\psi_1}/D{\psi_0}$.
Let us define a parameter $Y = (\psi_1/D\psi_0)$. This parameter is the 
ratio of the mean scattering strength of the thin screen and the distributed 
component. The ratio of the measured angular broadening ($\theta_H$) and the 
estimated value (i.e. $\theta_{\tau}$) is given by (Gwinn {\it et al.} 1993)
\begin{equation}
R_{\theta}=\frac{\theta_H}{\theta_{\tau}} = \left[ \frac{1 + 3(1-x)^2
\;Y}{1+6x(1-x)\;Y} \right]^{1/2}
\label{eq:ratio1}
\end{equation}
Depending on the validity of the assumption that the scattering material
is uniformly distributed along the line of sight, this ratio will deviate
from unity.

This relation has been derived and used by Gwinn {\it et al.} (1993) to 
infer values of $x$ along sight-lines to a few pulsars. However, they use
the available estimates of pulsar distance as an input in the analysis.
In the approach we wish to advance, we would like to invert this problem
to solve for the distance to a pulsar, given the distance to the 
discrete  scatterer. To be able to do so, we need to know the values for
$(\psi_1/D\psi_0)$ and the scatterer distance $d_s$ (=$xD$).
While in many cases it may be possible to identify a discrete scattering region
along a pulsar sight-line and use the known distance to such a region,
the contrast in the scattering rate $Y$ remains as one more
`unknown', unless this ratio can be assumed to deviate from unity by a large
factor. In the next section, we identify another similar but independent
relation which allows us to in fact also solve for the value of $Y$.

\begin{figure}
\epsfig{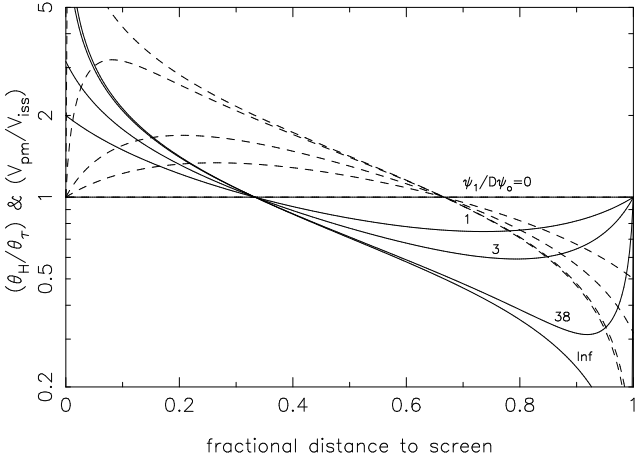}
\caption[]{The behaviour of the two ratios $R_{\theta}$ (solid line; eq. 6) 
and $R_v$ (dash line; eq. 10) as function of $x$ and $\psi_1/D\psi_0$.}
\label{fig:behaviour}
\end{figure}

\section{ Transverse Velocities}
\label{sec-vel}
The pulsar velocities inferred from the observed decorrelation time-scales of
interstellar scintillations compare well, on the average, with
those estimated from proper motion measurements (Cordes 1986; Gupta
{\it et al.} 1994). The reason for
possible
disagreements in the values estimated in these two ways is often
attributable to the breakdown of the assumption regarding the distribution
of the scattering medium along the sight-line. Gupta {\it et al.} (1994)
give the relevant expressions for the case of a single thin screen.
In this section, we
derive a general expression for this comparison, in terms of $x$, $D$, and
($\psi_1/D\psi_0$), for a two-component model.

The diffractive scintillation time scale associated with a screen at 
a location $x$ from the observer and having a characteristic irregularity 
size $a$, can be expressed as 


\begin{equation}
t_{\rm dif}(x) \;=\; \left[\frac{x\lambda D}{\pi a v_f}\right] = \left[
\frac{\lambda (1-x)}{\pi\theta v_{\rm pm}x} \right]
\end{equation}

\noindent
where $\lambda$ is the observing wavelength, $v_f$ 
is the apparent velocity of the scintillation pattern across the observer,
and $v_{\rm pm}$ is
the transverse velocity of the pulsar (which can be estimated from
proper motion measurements). 
It is easy to show, that 
the effective $t_{\rm dif}$ value corresponding to the distributed scattering
material is related to the harmonic mean of the $t_{\rm dif}^2$ values
for each of the sub-screens at different values of $x$. 
For our two-component model, the apparent $t_{\rm dif}$ can be 
expressed as

\begin{eqnarray}
\frac{1}{t_{\rm dif}^2} &=& \frac{\pi^2 v_{\rm pm}^2}{\lambda^2 D^2}
\int_0^Dx^2\;\psi(x)\;{\rm d}x \nonumber\\
 &=& \frac{\pi^2 v_{\rm pm}^2}{ 
\lambda^2}\frac{D\psi_0}{3}\;\left[1+3x^2Y \right]
\label{eq:tdiff}
\end{eqnarray}

With the knowledge of the measured values of $\tau_{\rm sc}$ and $t_{\rm
dif}$ the transverse velocity ($v_{\rm iss}$) of the pulsar can be estimated as (Cordes
1986; Gupta {\it et al.} 1994)
\begin{equation}
v_{\rm iss}^2 = \frac{Dc}{4\pi^2\tau_{\rm sc} t_{\rm dif}^2 \nu^2}
\end{equation}
\noindent
where $\nu$ is the observing frequency. With equations \ref{eq:unif} \&
\ref{eq:tdiff}, the above equation can be rewritten to define the ratio
$v_{\rm pm}/v_{\rm iss}$ as 


\begin{equation}
R_v = \frac{v_{\rm pm}}{v_{\rm iss}} = \left[ \frac{1 + 6x(1-x) Y}{1+3x^2 
Y} \right]^{1/2}
\label{eq:ratio2}
\end{equation}

In the above discussion, we have ignored the contribution to the
observed $t_{\rm dif}$ due
to the observer's motion and the possible motion associated with the
medium. However, the observed $t_{\rm dif}$ can be corrected for
these in cases where these contributions  are significant.
It is also worth mentioning that we have carefully examined the definitions of
the relevant observables and  the constants in all of the above equations.
Using detail simulations we have varified them to be consistent with those
used by Gupta et al. (1994).
\section{Estimation of Distances}
\label{sec-estdist}
Figure \ref{fig:behaviour} shows the behaviour of the
two ratios $R_{\theta}$ and
$R_v$ (equations \ref{eq:ratio1} \& \ref{eq:ratio2}) as a function of the
fractional distance $x$ to a discrete scatterer (from the observer) and 
the associated relative
strength of scattering $Y$. Solid lines are for the ratio
$R_{\theta}$, and the dash lines for $R_v$. Note that, except when 
$Y=0$, the two ratios deviate
from unity in general.
It is easy to see that, though the two ratios, $R_{\theta}$ \& $R_v$, show 
relative behaviour that is anti-symmetric around $x$=0.5, they do provide two 
independent relations between the quantities of interest. 
Hence, these relations can be used together 
to estimate the distance to a pulsar as well as the relative strength
of scattering $Y$
if the discrete scatterer distance ($d_s$=$xD$) is known.
\begin{figure}
\epsfig{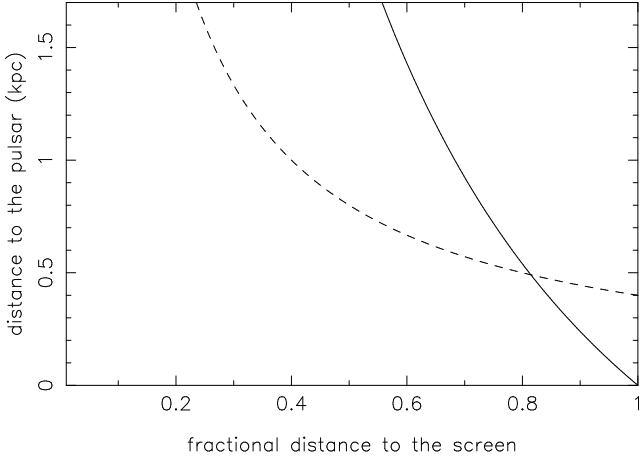}
\caption[]{Estimation of distance to Vela pulsar. Solid line gives the
relation between $x$ and $D$ according to equation \ref{eq:quadrat}
(with $r_{\theta}$ = 0.49 ${\rm kpc}^{-1/2}$ \& $r_v$ = 0.96 ${\rm kpc}^{-1/2}$), 
and the dash line is for
$D=(d_s/x)$ using $d_s$ = 400 pc. }
\label{fig:fig2}
\end{figure}


The four observable quantities necessary for this purpose are, 
($i$) the apparent angular
broadening $\theta_H$, ($ii$) the temporal broadening of the pulse 
profile $\tau_{\rm sc}$, ($iii$)
the diffractive scintillation time scale $t_{dif}$, and ($iv$) 
the proper motion $v_{\rm pm}$. 
While these depend on the pulsar distance, 
the two ratios $R_{\theta}$
\& $R_v$ (see eq.s \ref{eq:ratio1} \& \ref{eq:ratio2})
are independent of the pulsar distance.
However, in practice, the estimation of these two ratios based on the 
four observables
involves assumption of a distance, since $\theta_{\tau}\propto\sqrt{1/D}$,
$v_{\rm pm} = \mu\times D$, and $v_{\rm iss}\propto\sqrt{D}$, such that
both the ratios have $\sqrt{D}$ dependence. 

Let us therefore
express $R_{\theta}^2 = D\times r_{\theta}^2$, and $R_v^2 = D\times r_v^2$, where
$r_{\theta}$ \& $r_v$ can be treated as the estimated values of $R_{\theta}$ \& $R_v$ 
respectively if $D$ were to be equal to 1 kpc. 
Then, by using
equations \ref{eq:ratio1} \& \ref{eq:ratio2}, and eliminating 
$(\psi_1/D\psi_0)$, we get a useful relation between $D$ and $x$, as 
\begin{eqnarray}
D^2r_{\theta}^2r_v^2 (2x-3x^2) &+& \nonumber\\ 
Dr_v^2 (2x-1) &+& (3x-1)(x-1) \;=\; 0
\label{eq:quadrat}
\end{eqnarray}
Another independent relation, $d_s = xD$, would be obvious,
once the distance to the discrete scatterer is known. 
With these equations, the distance to the pulsar can be estimated,
once we know the distance to the thin screen scatterer and the ratios 
$r_{\theta}$,$r_v$. 

As we will discuss later in this paper, it may be possible to ascertain the
distance to a possible discrete scatterer in many cases.
However, as mentioned above, estimation of these two
ratios is possible only if all the four parameters are measured. Though the
measurement of $t_{\rm dif}$ and $\tau_{\rm sc}$ is more easy to come by, 
the measurement
of the other two quantities involves Very Long Baseline Interferometry and
therefore, the required measurements are available for only a few pulsars. 
For
instance, angular broadening measurements exist for only a handful of
pulsars (Gwinn {\it et al.} 1993). 
Reliable measurements of $\theta_H$ and $\mu$ are available so far
for only one pulsar, namely the Vela pulsar.
At 2.3 GHz, the measured values of the scattering 
parameters are, $\theta_H = 1.6\pm 0.2$ mas, $t_{\rm dif} = 15$ sec, and 
the decorrelation bandwidth is $68\pm 5$ kHz (Desai {\it et al.} 1992). 
The measured proper motion is $59.4\pm 2$ mas/yr (Bailes {\it et al.} 1990). 
Based on these, the corresponding values 
of $r_{\theta}$ \& $r_v$ are 0.49 kpc$^{-1/2}$ \& 0.96 kpc$^{-1/2}$, respectively.

\begin{figure}
\epsfig{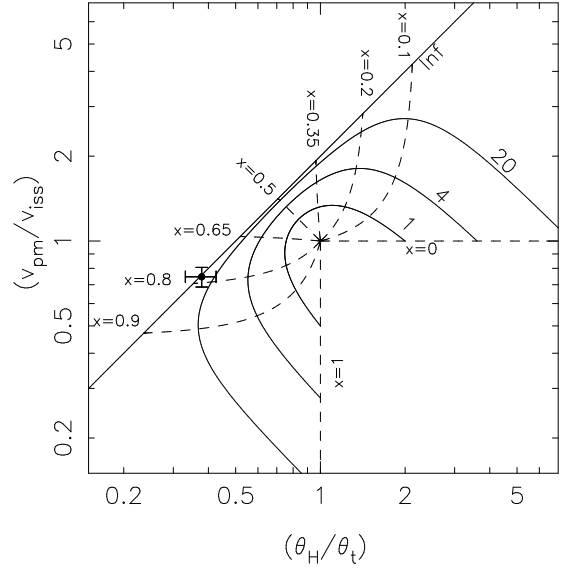}
\caption[]{A combined description of the dependences of the ratios 
$R_{\theta}$ and $R_v$ on $x$ and $Y$(=$\psi_1/D\psi_0$) corresponding to
the two-component
model for scattering. The solid \& the dashed curves show the relation 
between the two (velocity and the angular size) ratios for
different constant values of $Y$ and $x$ respectively.
The dot and the error bars correspond to the data on the Vela
pulsar. (See text for more details).}
\label{fig:ratios}
\end{figure}

The solid line in figure \ref{fig:fig2} shows the relation in 
equation \ref{eq:quadrat} after using these  ratios. The dashed
line in the figure corresponds to the relation $D=(d_s/x)$, where $d_s$, 
the distance
to the {\it discrete} scatterer 
is assumed to be equal to 400 pc. 
Although this is one of the estimates of the
estimates for the scatterer distance by Desai et al. (1992), our assumption 
is based on the following independent argument.
According to the Taylor \& Cordes
(1993) model for the electron density distribution, the Gum Nebula is
modelled at distance of about 500 pc with a radius of about 180 pc.
Thus, the mean distance of the section of the Gum nebula that may be in the 
foreground of the Vela pulsar would be in the range 300-500 pc. 
Hence, we consider the relevant mean distance to be nominally 400 pc.  
 The intersection of 
these two curves in figure \ref{fig:fig2}
represents the solution in terms of $x$ \& $D$.
In the present case, the intersection is at $x=0.8$ and $D=500$ pc, 
which agrees with the generally accepted distance
to the Vela pulsar. 
As can be seen from the solid-line curve in Fig. 2, that an uncertainty of 
the order of 100 pc in the scatterer distance will imply very small change 
in the fractional distance ($x$). 
So the distance to the Vela pulsar would be $500\pm 125$ pc
considering the worst case error.
As we shall see in a following section, an independent estimation of 
the distance the Vela pulsar (see Table 1) also supports our assumtion 
of the scatterer distance, remembering that the fractional distance is about 0.8.

It is worth recalling here that Desai et al. (1992) also derive an effective 
fractional distance of $0.81\pm 0.03$, by however assuming a 500 pc distance to
the Vela pulsar (Frail \& Weisberg 1990), implying 
an effective distance of 400 pc to the scatterer. However, Desai et al. argue that,
since the near-edge of the Gum Nebula is at about 270 pc (Sivan 1974; Reynolds
1976), the scattering observed in Vela pulsar cannot be due to the Gum nebula. When
they consider a uniformly distributed scattering component with strength 
close to the
Galactic disk (Cordes et al. 1985), then the fractional distance to the scatterer 
turns out to be $x=0.87$. Alternatively, they argue that, if the Gum nebula 
scatters intrinsically as 
much as 5\% as sstrongly as the other scattering screen, then the screen is pushed
to a fractional distance of $x=0.95$. However, we have assume an effective 
fractional distance of 400 pc as the distance to the scatterer 
and, by our independent method, we find a solution the $x$ to be equal to $0.8$,
implying the distance to the Vela pulsar as 500 pc. 
In the framework of our present two-component model, our results are in 
excellent agreement with those by Desai et al. (1992). 


\section{Description in the $(v_{pm}/v_{iss})$--$(\theta_H/\theta_{\tau})$
plane}
\label{sec-probing}

Although, in practice, the above procedure is convenient to use, 
it is instructive  
to express the dependences of the two ratios $R_{\theta}$ \& $R_v$ on $x$ \& 
$\psi_1/D\psi_0$
in a general form as shown in figure 
\ref{fig:ratios}. The two sets of curves indicated by the continuous and the
broken (dashed) lines correspond to different constant values 
of $Y\;=\;(\psi_1/D\psi_0)$ and $x$ respectively.
The dot (with the error bars) represents the data on the Vela pulsar using the above
derived distance.
The relative scattering strength $Y$ 
of the discrete scatterer, as seen from 
the diagram, is very high as would be expected from the known 
anomalous scattering in this direction.

Given the measurements of the four required observables, the possible
combinations of $D$, $x$ and $Y$ can be represented on this 
log-log plot by a straight line of $45^0$ slope (parallel to the line 
corresponding to $Y=\infty$.
Each point on such a line will correspond to a distinct value of the pulsar 
distance (depending on $r_{\theta},r_v$), and in general, would imply a
unique combination of $x$ \& $Y$.
The {\it correct} distance would correspond to one such point, for which the 
implied value of $x$ is consistent with our knowledge of $d_s$.

There are some {\it prohibited areas} in this diagram, namely, the lower-right
square ( $R_{\theta} > 1 > R_v$) and the upper-left triangle ($r_v/r_{\theta} > 2$). 
For example, the range of distances corresponding to the sections through the 
lower-right square can be rejected. When such a rejectable range is large, 
it would imply that the discrete scatterer is dominant and that it is 
quite close to either the pulsar or the observer. In such cases,
some other rather simple considerations may be enough to yield useful 
limits on the pulsar distance.
On the other hand, the `impossibility'  of $r_v/r_{\theta}>2$ in the
present two-component model, may be 
utilized to constrain the values of some of the observables 
that may not be known or have large uncertainties.
Similar considerations 
would  also apply  for the other `prohibited' square region.

In certain cases, when it is possible to assume absence of any
discrete component of enhanced scattering along a sight-line, the region of
interest is confined to the central part of the diagram that is bounded
by, say, $Y$ = 0.5. This makes it possible then to estimate
the pulsar distance within an uncertainty of 20\% or so, even when only
one of the two ratios, $r_{\theta}$ or $r_v$, may be known. It is worth
remembering that in such cases the parameter $x$ is not important.

In certain other cases, it may be clear that a discrete screen is the
dominant scatterer (e.g. like the Gum nebula in the case of the 
Vela pulsar). Then, the relation in equation \ref{eq:quadrat} 
can be approximated
to assume simple forms where the ratios $r_{\theta}$ and $r_v$ can be
decoupled.  The two resultant relations can then be expressed in terms
of $d_s$, as
\begin{eqnarray}
x\;=\;1\;-\;2d_s r_v^2 \;\;\;\;\; or \;\;\;\;\;
x\;=\;1\;-\;{\frac{1}{2}}d_s r_{\theta}^2
\label{eq:linear}
\end{eqnarray}
 In this regime, the fractional distance $x$ and the pulsar distance
can be estimated reasonably accurately 
even when only one of the ratios is known, provided the `true' value of
$x$ is not very small (i.e. not $<$0.1).  
Otherwise, both the ratios need to be
known so as to avoid a large error in the distance estimation.
In general, the low-low and the high-high combinations of the $R_{\theta}$,$R_v$
values (as also when ($R_{\theta}/R_v$) is nearly equal to 2) definitely
indicate the presence of a dominant discrete scatterer. Therefore, when
the presence of a dominant discrete scatterer can be assumed, this 
connection between $R_{\theta}$ \& $R_v$ can
be exploited suitably when only one of two ratios is known, and the other
ratio is required to be estimated.   

It is easy to note from the diagram (as also from eq. \ref{eq:quadrat})
that the distance estimation becomes independent of one of the two
ratios ($r_{\theta}$ or $r_v$) when $x$= $0,\;(1/3),\;(2/3)$ or
$1$, while that ratio will be important only for the determination of 
$Y$. A nearly similar behaviour is also evident when 
$Y>>1$.
For example, the distance determination would be less sensitive
to  uncertainties in $R_{\theta}$ ($R_v$) when $x>(2/3)$ ($x<(1/3)$).
This should dictate the choice of the measurement and the 
corresponding relation to be used
(of the two in equation \ref{eq:linear}, for example) for determining
$x$ and $D$.

%

This diagram thus provides us a useful and a nearly complete description 
in terms of the dependence of the two observable ratios on the two
model parameters $x$ \& $Y$, which represent the 
fractional distance and the relative strength of scattering respectively
for the discrete scatterer in the assumed two-component model.

\section{Anomalous scattering \& distance to PSR B0736--40}
\label{sec-distance}
In some cases, a dominant discrete scatterer may be common to sight-lines
to a number of pulsars. In such a case, characterization of the scattering
region can be attempted using the data on the four observables which may
be available for only a few pulsars and the results could be used in
determining distances to the other pulsars with nearby sight-lines.
This relaxes the requirement that all four observables be available
for the rest of the pulsars in the set and, in fact, the knowledge of DM
and temporal scatter broadening  alone suffices for distance
determination.  We illustrate this possibility
by considering the Gum nebula region as a common scatterer for a number of
pulsar sight-lines. 

Taylor \& Cordes (1993) do include the enhanced electron
density in the Gum nebula region in their model, but state
that the modelling of this component 
is far from complete since a proper modelling would require
many more constraints (for both the dispersion and the scattering) 
than are presently available.  They have, therefore, assumed
the fluctuation parameter (which quantifies the amount of scattering in the
medium, given the value for the electron density) associated with the
Gum nebula to be equal to {\it zero}. In the following part of this section,
we show that the data on the Vela pulsar can be exploited to {\it calibrate}
this fluctuation parameter (as defined in the Taylor \& Cordes model).

 In a simple exercise, we varied the assumed number density and the
fluctuation parameter associated with the Gum nebula in the 
framework of the Taylor \& Cordes model to obtain values of these 
parameters that would be consistent with the known DM (69 pc cm$^{-3}$) 
\& $\tau_{sc}$ (8.25 ms at 327 MHz)  as well as the derived
distance (500 pc) to the Vela pulsar.
As shown in Table \ref{table:table1} (columns 1, 4 \& 5),
the electron density in the Gum nebula region needs to be 
about $60$\% higher than that suggested in the
Taylor \& Cordes model.  The value of the fluctuation parameter is
about $6.3$, quite close to that for
the spiral arm component in the Taylor \& Cordes model.
Although the Gum nebula is quite complicated in its structure
and morphology, it may not be unreasonable to assume that the above 
estimates would be more or less valid for other sight-lines through the Gum 
region. 

With this view, we consider another pulsar B0736--40 ($l=254^{\circ}.2$, 
$b=-9^{\circ}.2$ \& DM = 160.8 pc cm$^{-3}$), for which the estimated 
distance on the basis of the Taylor \& Cordes (1993) model is $>11$ kpc
(and therefore would be placed beyond the `outer' spiral arm). 
As part of our survey (in 1996) to measure the temporal broadening of pulse 
profiles at 327 MHz (Ramachandran {\it et al.} 1997), we observed
this object and measured the
temporal broadening to be $76\pm 3$ ms. Assuming   
the excess scattering is due to the Gum nebula,
we again seek a combination of the electron density and the fluctuation 
parameter that would be consistent with the DM \& $\tau_{sc}$ values, and the
distance estimate associated with it. As can be seen from 
Table \ref{table:table1}, both the cases, the Vela pulsar
\& B0736--40, are consistent with the electron density  and the fluctuation 
parameter values of 0.32 cm$^{-3}$ \& $\sim 6.3$ respectively.
This, therefore, allows us to  derive a distance 
to B0736--40  as $\sim 4.5$ kpc (with an uncertainty of about 0.8 kpc) 
based on our measurement of the temporal broadening \& the 
{\it calibration} from the Vela pulsar data.

We also find that the excess temporal broadening cannot be accounted for
by even a large increase in the electron density associated with the
`outer' spiral arm component. 
The z-height of about 700 pc, based on our new estimate, implies that
the `true' distance could be even shorter given the possibility that
the effective scale-height of the electron distribution  
may be somewhat under-estimated in the Taylor \& Cordes model.
As one of the important implications of this distance determination, the 
estimated velocity of the pulsar would now be less than 1600 km/sec
(using a proper motion of about 72.5 mas/yr)
rather than a value $>$3780 km/sec as was implied by the earlier distance
estimate. 

\begin{center}
\begin{table}
\begin{tabular}{c|c|c|c|c} \hline
{\bf $n$} & \multicolumn{2}{|c|}{\bf 0736--40} & \multicolumn{2}{|c|}{\bf
Vela pulsar} \\ \cline{2-3}\cline{4-5}
$\times 0.2$ cm$^{-3}$ & {\bf f} & {\bf D} (kpc) & {\bf f} & {\bf D} (kpc) \\ \hline\hline
1.0   &  15.3  &  11.0  &  7.3   &  0.61 \\
1.1   &  12.8  &  11.0  &  7.0   &  0.59 \\
1.2   &  10.6  &   9.1  &  6.85  &  0.57 \\
1.3   &  8.75  &   7.1  &  6.75  &  0.55 \\
1.4   &  7.45  &   5.9  &  6.6   &  0.53 \\
1.5   &  6.40  &   5.1  &  6.5   &  0.52 \\
1.6   &  5.65  &   4.4  &  6.4   &  0.51 \\
1.7   &  5.00  &   3.8  &  6.3   &  0.50 \\
1.8   &  4.45  &   3.3  &  6.3   &  0.49 \\
1.9   &  4.00  &   2.9  &  6.2   &  0.48 \\
2.0   &  3.65  &   2.5  &  6.2   &  0.48 \\ \hline
\end{tabular}
\caption[]{Distances (for B0736--40 \& the Vela pulsar) estimated by 
changing the electron density for the Gum nebula component in the 
model of Taylor \& Cordes (1993), and the corresponding 
fluctuation parameter implied by the observed temporal broadening.
The first column gives the `enhancement factor' for the assumed 
electron density, 
the second \& fourth columns give the values of the fluctuation parameter
and the third \& the fifth columns give the corresponding distances in
the two cases.}
\label{table:table1}
\end{table}
\end{center}

Any more detailed modelling of the distribution of electron density and 
its fluctuation across the Gum nebula region is beyond the scope of 
this paper, and needs new 
measurements of scattering parameters of a number of pulsars that could
be behind this region.
We, however, have shown how distance estimates for many such pulsars 
could be refined
already with the help of the scatter broadening measurements.

\section{Discussion and Conclusions}
\label{sec-discussion}
We have described in this paper a method to study effectively the
distribution of the electron density fluctuations in the interstellar
medium by assuming a simple model that seems consistent in most
cases.  
Essentially, suitable combinations of four measurable 
quantities for pulsars, namely
the temporal broadening of pulse profiles, angular broadening of the
source, diffractive scintillation time scale, and the proper motion, 
allow us to solve for the parameters of the scatterer as well as for
the pulsar distance.
The important simplifying assumption that has gone into this
analysis is that the scattering along the sight-line can be adequately
modelled in terms of just two components, a thin screen scatterer and 
a uniformly distributed component. 
It is clear that for lines of sight which pass
through, say, two spiral arms, this simple picture would need to be modified.
However, for a large number of pulsars, our present assumption can be justified
(Gwinn {\it et al.} 1993). One would  naturally
attempt to model the local medium first and use that knowledge while
probing progressively the farther section.
In such a case, it is easy to see that the number of {\it unknowns} in the 
problem at any stage would not be too many to handle.

One of the important ingredients in our method is the knowledge of the 
distance to the discrete scatterer,  particularly if such a 
scatterer is the dominant source of scattering.  
For most nearby pulsars (within 2 kpc or so) such a scatterer, if any,
should be identifiable as an HII region or the Stromgren region of some 
OB stars or as a region associated with a supernova remnant. It is 
therefore reasonable
to assume that, in most such cases, the scatterer distance would be available
(see, for example, Prentice \& ter Haar, 1969). For pulsars at 
high galactic latitudes, it may be possible to assume that the dominant
scatterer is at a z-height defined by the scale-height of the HII region
distribution in the disk of the galaxy. For somewhat distant pulsars 
well within the disk, the discrete scatterer may be identifiable with a spiral
arm component. (In such cases, the distance limits that may be available from HI
absorption measurements can be incorporated in the analysis as additional
constraints.)
We therefore do not consider the scatterer distance as a difficult
ingredient to supply.

The method we have proposed can therefore be used for reliable determination 
of  distances to a large number of pulsars
as well as for probing a usefully large volume of the galaxy for its
electron density distribution. A systematic and intensive observational
program to measure the scattering effects and proper motions of pulsars
would be extremely fruitful. In cases with dominant discrete scatterers,
it may enough to measure quantities related to only one of the two ratios.
Since the data on temporal broadening are already available for a number of
pulsars, extending the VLBI measurements of angular broadening (e.g.
Gwinn {\it et al.}, 1993) to a corresponding set of pulsars will be
worthwhile.
Such measurements,
unlike the proper motion measurements, do not need long time-baselines
or phase-referencing.

Our simple-minded analysis related to the Gum nebula region illustrates
how even a less elaborate characterization of such extended scatterers
using a few calibrators is useful in estimating the distance to pulsars
with limited data on scattering. The present example also shows how 
such information can be used in the framework of the Taylor \& Cordes model.
The particular case of B0736--40 amply emphasizes the importance of
reliable distances in  the estimation of pulsar velocities, and we would like to
stress that there could be many such cases, particularly amongst 
the high galactic latitude pulsars.

To conclude, the method presented here suggests an attractive possibility
for distance estimation based on observables related to the interstellar 
scattering and scintillations. Such estimations would then play an important
role in refining the present model for the distribution of electrons in
our galaxy.

\section*{Acknowledgement}
We would like to thank Rajaram Nityananda, V. Radhakrishnan \& J. H.
Seiradakis  for critical reading of the 
manuscript and many useful suggestions for refining the text. We also
thank an anonymous referee for his critical comments.

\end{document}